\documentclass[twocolumn]{aastex631}

\usepackage{graphicx}
\usepackage{amssymb,amsmath}
\usepackage{aas_macros}
\usepackage[T1]{fontenc}
\usepackage{bm}
\usepackage{xcolor}
\usepackage{soul} %% for \sout

\newcommand{\dd}{\ensuremath{\mathrm{d}}}%
\newcommand{\ee}[0]{\ensuremath{\mathrm{e}}}
\newcommand{\vct}[1]{\ensuremath{\bm{#1}}}%

\newcommand{\Lmax}{\ensuremath{L_{\text{max}}}}%
\newcommand{\Lmin}{\ensuremath{L_{\text{min}}}}%
\newcommand{\Lc}{\ensuremath{L_{\text{c}}}}%
\newcommand{\rg}{\ensuremath{r_{\text{g}}}}%
\newcommand{\tauc}{\ensuremath{\tau_{\text{c}}}}%
\renewcommand{\FL}{\mathrm{FL}}

\begin{document}

\title{Diffusion of relativistic charged particles and field lines in isotropic turbulence: II. Analytical models}

\author[0000-0002-0454-6823]{Marco Kuhlen}
\affiliation{Institute for Theoretical Particle Physics and Cosmology (TTK), RWTH Aachen University, 52056 Aachen, Germany}
\email{marco.kuhlen@rwth-aachen.de}

\author[0000-0002-2197-3421]{Philipp Mertsch}
\affiliation{Institute for Theoretical Particle Physics and Cosmology (TTK), RWTH Aachen University, 52056 Aachen, Germany}
\correspondingauthor{Philipp Mertsch} 
\email{pmertsch@physik.rwth-aachen.de}

\author[0000-0002-5611-095X]{Vo Hong Minh Phan}
\affiliation{Sorbonne Université, Observatoire de Paris, PSL Research University, LERMA, CNRS UMR 8112, 75005 Paris, France}
\affiliation{Institute for Theoretical Particle Physics and Cosmology (TTK), RWTH Aachen University, 52056 Aachen, Germany}
\email{vhmphan@physik.rwth-aachen.de}

\preprint{TTK-22-36}

\date{\today}

\begin{abstract}
The transport of high-energy particles in the presence of small-scale, turbulent magnetic fields is a long-standing issue in astrophysics. 
Analytical theories on transport perpendicular to the large-scale magnetic field disagree with numerical simulations at rigidities where the particles' gyroradii are slightly smaller than the correlation length of turbulence. 
At the same time, extending the numerical simulations to lower rigidities has proven computationally prohibitive. 
We present an analytical model for the perpendicular transport, based on (1) initial particle transport along field lines, (2) the transport of field lines and (3) the eventual decorrelation of particles from field lines. 
Transport parallel to the large-scale field is governed by pitch-angle scattering and so for times larger than the inverse pitch-angle diffusion coefficient, particles spatially diffuse in the parallel direction. 
Our results suggest that perpendicular diffusion occurs when particles have displaced in the perpendicular direction by a few correlation lengths of turbulence. 
We have tested the analytical theory by running a large suite of test particle simulations at unprecedentedly low rigidities, making extensive use of graphical processing units (GPUs). 
Our numerical results exhibit a non-standard rigidity-dependence for the perpendicular diffusion coefficient at intermediate rigidities. At the lowest rigidities, the standard rigidity-dependence is recovered. The simulated diffusion coefficients are nicely reproduced by our analytical model. We have traced the non-standard rigidity-dependence to a subdiffusive phase in the field line transport. Our study confirms our understanding of the escape of cosmic rays from the Galactic halo and its rigidity-dependence. It also rejects speculations about different rigidity-dependencies of the parallel and perpendicular diffusion coefficients at very low rigidities that were invoked to explain non-standard gamma-ray spectra from the Galactic Centre. Other possible applications concern particle acceleration at perpendicular shocks. 
\end{abstract}

%\maketitle

% ----------------------------------------------------------------------------------------
% ----------------------------------------------------------------------------------------
% ----------------------------------------------------------------------------------------
\section{Introduction}

Much of the universe is pervaded by magnetic fields, oftentimes turbulent, and high-energy charged particles, cosmic rays (CRs). The transport of charged particles in the presence of turbulent magnetic fields is a fundamental question in astrophysics. The presence of a large-scale, coherent magnetic field breaks the isotropy and leads to the differentiation of particle transport along and across this large-scale field. Transport along the field has been understood a while ago to be the result of initially ballistic gyrations, followed by pitch-angle scattering due to resonant interactions~\citep{1966ApJ...146..480J,1966PhFl....9.2377K,1967PhFl...10.2620H}. Eventually, pitch-angle scattering results in spatial diffusion along the large-scale magnetic field~\citep{1970ApJ...162.1049H}. Perpendicular transport has proven much more difficult to model, but is generally believed to be the combination of transport along field lines and transport of the field lines~\citep{2009ASSL..362.....S,2020SSRv..216...23S}.

Analytical models can be compared to so-called test particle simulations. Those are based on solving the equations of motion of individual particles in a turbulent magnetic field simulated on a computer. (See \citealt{2020Ap&SS.365..135M} for a review.) At low reduced rigidities $\rg / \Lc$ (where $\rg$ is the gyroradius and $\Lc$ is the correlation length of turbulence,~\citealt{2002JHEP...03..045H}), particle trajectories have to be followed for very long times before particle motion becomes diffusive. Previously, simulations have only been possible at relatively large $\rg/\Lc \gtrsim 10^{-2}$~\citep{1999ApJ...520..204G,2001PhRvD..65b3002C,2002GeoRL..29.1048Q,2002ApJ...578L.117Q,2007JCAP...06..027D,2012JCAP...07..031G,2016MNRAS.457.3975S,2016MNRAS.459.3395P,2017ApJ...837..140S,2018JCAP...07..051G,2020ApJ...889..123S,2020MNRAS.498.5051R,2022SNAS....4...15R}. For parameters typical of the interstellar medium, e.g.\ an rms field strength of $4 \, \mu \text{G}$ and $\Lc = 30 \, \text{pc}$, this corresponds to rigidities $\gtrsim 1 \, \text{PV}$. 

Recently, it has become apparent that none of the analytical theories developed over the last decades, e.g.~\citep{2003ApJ...590L..53M,2010ApJ...720L.127S}, can describe the results of test particle simulations for isotropic turbulence~\citep{2020PhRvD.102j3016D}. For instance, these theories typically predict a rather weak, but falling dependence of the ratio of perpendicular and parallel mean free paths, $\lambda_{\perp}/\lambda_{\parallel}$ on $\rg/\Lc$. Instead, in simulations, the ratio increases for $\rg/\Lc \lesssim 1$ and drops sharply for $\rg/\Lc \gg 1$~\citep{2007JCAP...06..027D,2016MNRAS.457.3975S,2020PhRvD.102j3016D}. This failure of analytical theories is most worrisome and casts doubt on our general understanding of perpendicular transport. Furthermore, if the difference in the rigidity-scaling of $\lambda_{\parallel}$ and $\lambda_{\perp}$ were confirmed, this would have important consequences for phenomenological applications of such models, for instance in modelling of Galactic cosmic rays and diffuse emission~\citep{2012PhRvL.108u1102E,2017JCAP...10..019C,2018JCAP...07..051G,2022SNAS....4...15R}. 

% ----------------------------------------------------------------------------------------
% ----------------------------------------------------------------------------------------
% ----------------------------------------------------------------------------------------
\section{Method}

% ----------------------------------------------------------------------------------------
% ----------------------------------------------------------------------------------------
\subsection{Test particle simulations}

We have run a large suite of test particle simulations with a field composed of a large-scale, coherent magnetic field, taken to point in the $z$-direction and a small-scale, static, turbulent magnetic field $\vct{\delta B}(\vct{r})$. (The magneto-static approximation is valid for particle velocities exceeding the typical speed of the turbulent waves, e.g.\ the Alfv\'en speed.) The turbulent modes are isotropically distributed and follow a Kolmogorov~\citep{1941DoSSR..30..301K} power spectrum $P(k) \propto k^{-5/3}$ for $k > (2 \pi) / \Lmax$ with an outer scale $\Lmax$, where $\langle \vct{\delta \tilde{B}}(\vct{k}) \vct{\delta \tilde{B}}^*(\vct{k}') \rangle \propto P(k) \delta^{(3)}(\vct{k} - \vct{k}')$. Here, angled brackets denote averages over an ensemble of turbulent magnetic field. The correlation length $\Lc \simeq \Lmax / 5$ and the overall turbulent strength is parametrised by $\eta \equiv \delta B^2 / (B_0^2 + \delta B^2)$ with $\delta B^2 \equiv 8 \pi \int_0^{\infty} \dd k \, k^2 P(k)$.

From the particle trajectories, $\{ x(t), y(t), z(t) \}$, we have computed running parallel and perpendicular diffusion coefficients~\citep{2009ASSL..362.....S},
\begin{align}
d_{\parallel}(t) &\equiv \frac{1}{2} \frac{\dd}{\dd t} \langle (\Delta z(t))^2 \rangle \, , \label{eqn:def_dpar} \\
d_{\perp}(t) &\equiv \frac{1}{4} \frac{\dd}{\dd t} \left( \langle (\Delta x(t))^2 \rangle + \langle (\Delta y(t))^2 \rangle \right) \, ,
\label{eqn:def_dperp}
\end{align}
with $\Delta x(t) = x(t) - x(0)$ etc.
Here, the angled brackets denote the average over initial particle directions and realisations of the turbulent magnetic field.

We have also computed the magnetic field lines $\{x^\FL(s), y^\FL(s), z^\FL(s) \}$, parametrised by their arc-length $s$, by solving the field line equations for individual realisations of $\vct{\delta B}$. Note that the coordinates of the field lines are denoted with a superscript $\FL$ for ``field line''. From this, we can compute the running field line diffusion coefficient~\citep{2009ASSL..362.....S}, 
\begin{equation}
d_{\text{FL}}(s) \equiv \frac{1}{4} \frac{\dd}{\dd s} \left( \langle (\Delta x^\FL(s))^2 \rangle + \langle (\Delta y^\FL(s))^2 \rangle \right) \, . \label{eqn:def_dFL_z}
\end{equation}
Note also that in axisymmetric turbulence, $\langle (\Delta x^\FL(s))^2 \rangle = \langle (\Delta y^\FL(s))^2 \rangle$. 

To cover the large dynamical range of $\Lmax/\Lmin \simeq 10^{6}$ required for resonant scattering at low rigidities, we have employed the nested grid method~\citep{2012JCAP...07..031G,2018JCAP...07..051G,2020Ap&SS.365..135M} for setting up a large enough number of individual realisations of $\vct{\delta B}$ on a computer. We employ 16 levels of subgrids and the largest grid extends over $40 \, \Lc$ so as to minimise periodicity artefacts. We have made use of graphical processing units (GPUs) that allow for a significant speed up~\citep{2016NewA...45....1T,2022ApJ...927..110K}. In a companion paper~\citep{Kuhlen:2022tov}, we provide a detailed discussion of the method and its possible systematic errors as well as the time-dependent running diffusion coefficients.

% ----------------------------------------------------------------------------------------
% ----------------------------------------------------------------------------------------
\subsection{Analytical model}

Our analytical model is based on the same principles as earlier works~\citep{1997ApJ...485..655B,2003ApJ...590L..53M,2002ApJ...578L.117Q,2004ApJ...616..617S,2019ApJ...881L..27S}, but with a crucial difference. (1) Initially, particles follow magnetic field lines, at first ballistically, later diffusively. (2) The transport of field lines is also transitioning from an early ballistic phase to an eventual diffusive regime. At variance with earlier studies, however, we allow for the possibility of a subdiffusive episode in field line transport where $\langle (\Delta x)^2 \rangle \propto s^{\alpha}$ with $\alpha < 1$. (3) Eventually, as the particle displacement in the perpendicular direction exceeds about a correlation length, particles can be considered decoupled from individual field lines and start to diffuse~\citep{2019ApJ...881L..27S}. 
Based on these principles, we can express the running perpendicular diffusion coefficient $d_{\perp}(t)$ in terms of the field line diffusion coefficient, $d_{\text{FL}}(z)$, the particle running parallel diffusion coefficient $d_{\parallel}(t)$ and the particle mean-square parallel displacement, $\langle (\Delta z(t))^2 \rangle$,
\begin{align}
d_{\perp}(t) &= \frac{\dd}{\dd t}  \int_0^{\sqrt{\langle (\Delta z(t))^2 \rangle}}  \dd z' d_{\text{FL}}(z') \nonumber \\
&= \frac{d_{\text{FL}}(\sqrt{\langle (\Delta z(t))^2 \rangle})}{\sqrt{\langle (\Delta z(t))^2 \rangle }} d_{\parallel}(t) \, . \label{eqn:composition_dperp}
\end{align}
The particle mean-square parallel displacement, $\langle (\Delta z(t))^2 \rangle$ follows from integrating eq.~\eqref{eqn:def_dpar}, 
\begin{equation}
\langle (\Delta z(t))^2 \rangle = 2 \int_0^t \dd t' d_{\parallel}(t) \, .
\end{equation}
Eq.~\eqref{eqn:composition_dperp} represents the main result of this paper. 
In the following material, we will discuss briefly an heuristic derivation for this equation. A more detailed derivation based on microscopic treatments is also presented in Appendix \ref{appendix:micro}.

% ----------------------------------------------------------------------------------------
\subsubsection{Heuristic derivation}

To derive eq.~\eqref{eqn:composition_dperp}, we employ first the definition of the running FL diffusion coefficient, eq.~\eqref{eqn:def_dFL_z}.
Integrating eq.~\eqref{eqn:def_dFL_z},
\begin{equation}
\langle (\Delta x^{\text{FL}}(z))^2 \rangle = 2 \int_0^z \dd z' \, d_{\text{FL}}(z') \, , \label{eqn:rperp2}
\end{equation} 
where $\Delta x^\FL(z)=x^\FL(z)-x^\FL(0)$. Now, we assume that the vertical displacement of the particles before perpendicular diffusion taking place is only due to the perpendicular displacement of the field line which means  
\begin{align}
\langle(\Delta x(t))^2\rangle & = \langle \Delta x^{\text{FL}}(s=\langle \Delta z^2(t) \rangle)\rangle,  \label{eq:attach}\\ 
& = 2 \int_0^{\sqrt{\langle \Delta z^2(t) \rangle}} \dd z' \, d_{\text{FL}}(z'). \label{eqn:rperp3}
\end{align}
Substituting eq.~\eqref{eqn:rperp3} into eq.~\eqref{eqn:def_dperp} then leads to eq.~\eqref{eqn:composition_dperp}.

% ----------------------------------------------------------------------------------------
\subsubsection{ODE approach for running field line diffusion coefficient}

The running field line diffusion coefficient $d_{\text{FL}}$ can be approximated as the solution of a set of coupled ordinary differential equations (ODEs)~\citep{2016ApJS..225...20S}. 
Here, we consider a simplified setup where the dispersion $\sigma_z^2$ in the field line direction is ignored. 
In addition, we approximate the arc-length along the field line with the coordinate $z$ along the background field direction which should be a good approximation for small turbulence levels $\eta$. 

We start by integrating the field line equation for the $x$-component,
\begin{equation}
\langle (\Delta x^{\text{FL}}(z))^2 \rangle = \frac{1}{B_0^2} \int_0^z \dd z' \int_0^z \dd z'' \, \langle \delta B_x(z') \delta B_x(z'') \rangle.
\end{equation}
We can then employ the properties of homogeneous turbulence meaning $\langle \delta B_x(z') \delta B_x(z'') \rangle = \langle \delta B_x(z'-z'') \delta B_x(0) \rangle $ to show that
\begin{equation}
\langle (\Delta x^{\text{FL}}(z))^2 \rangle = \frac{2}{B_0^2} \int_0^z \dd z' \, (z-z') \langle \delta B_x(z') \delta B_x(0)\rangle. 
\end{equation}
The running field line diffusion coefficient is then
\begin{align}
    d_\FL &=\frac{1}{2}\frac{\dd \langle (\Delta x^{\text{FL}}(z))^2 \rangle}{\dd z}, \label{eq:d_FL_couple}\\
    &=\frac{1}{B_0^2} \int_0^z \dd z' \langle \delta B_x(z') \delta B_x(0)\rangle.
\end{align}
We can now differentiate the above equation with respect to $z$ to obtain
\begin{widetext}
\begin{align}
    \frac{\dd d_\FL}{\dd z} &=\frac{\langle \delta B_x(z) \delta B_x(0)\rangle}{B_0^2},\\
    &=\frac{1}{B_0^2}\int\dd^3 \vct{k}\int \dd^3\vct{k}' \langle \delta \tilde{B}_x(\vct{k})\delta\tilde{B}^*_x(\vct{k}')\rangle\left\langle {\rm e}^{i\vct{k}\cdot\vct{x}(z)-i\vct{k}'\cdot\vct{x}(0)}\right\rangle,\\
    &=\frac{1}{B_0^2}\int\dd^3 \vct{k} P(k)\cos(k_\parallel z)\left\langle {\rm e}^{i\vct{k}_\perp\cdot\left(\vct{x}_\perp(z)-\vct{x}_\perp(0)\right)}\right\rangle. \label{eq:d_FL-derive}
\end{align}
\end{widetext}
where we have made use of Corrsin's independence hypothesis~\citep{1959AdGeo...6..161C} and assumed independence of the characteristic functions in the parallel and perpendicular directions. For an isotropic Kolmogorov spectrum,
\begin{equation}
P(k) \propto \left( \delta_{ij} - \frac{k_i k_j}{k^2} \right) \left( \frac{k}{k_0} \right)^{-11/3} \, . \label{eq:redef-power} 
\end{equation}
for $k > k_0$ and zero otherwise.
We assume also  a Gaussian distribution of the field lines in the perpendicular direction such that 
\begin{equation}
\left\langle {\rm e}^{i\vct{k}_\perp\cdot\left(\vct{x}_\perp(z)-\vct{x}_\perp(0)\right)}\right\rangle = {\rm e}^{-\frac{1}{2}k_\perp^2\langle(\Delta x^\FL(z))^2\rangle}. \label{eq:FL-gaussian}
\end{equation}
Substituting eq.~\eqref{eq:redef-power} and eq.~\eqref{eq:FL-gaussian} into eq.~\eqref{eq:d_FL-derive} and, together with eq.~\eqref{eq:d_FL_couple}, we have a system of coupled ODEs to estimate the running field line diffusion coefficient,
\begin{align}
d_{\text{FL}} &= \frac{1}{2} \frac{\dd \langle (\Delta x^{\text{FL}}(z))^2 \rangle}{\dd z} \, , \label{eqn:ODE_system_1} \\
\frac{\dd d_{\text{FL}}}{\dd z} &= \frac{2 \pi}{\mathcal{N}} \frac{\delta B^2}{B_0^2} \int_{-1}^{1} \dd(\cos\theta) \, \int_{k_0}^{\infty} \dd k \, k^2 \left( \frac{k}{k_0} \right)^{-11/3} \nonumber \\
& \mkern-36mu \times (1 + \cos^2 \theta) \ee^{-\frac{k^2}{2} \langle (\Delta x^{\text{FL}}(z))^2 \rangle \sin^2 \theta} \cos (k \cos \theta z) , \label{eqn:ODE_system_2}
\end{align}
Here, the normalisation constant $\mathcal{N}$ is defined as  
\begin{equation}
\mathcal{N} = 8 \pi \int_{k_0}^{\infty} \dd k \, k^2 \left( \frac{k}{k_0} \right)^{-11/3} 
\end{equation}

\begin{figure}[tbh]
\includegraphics[scale=1]{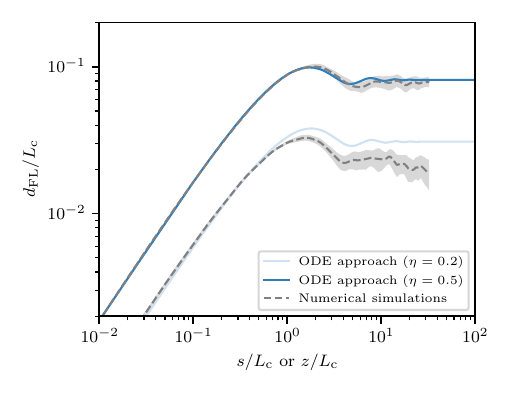}
\caption{Running field line diffusion coefficient $d_{\text{FL}}$. For turbulence levels of $\eta = 0.2$ and $0.5$ we compare the result of the numerically integration of eqs.~\eqref{eqn:ODE_system_1} and \eqref{eqn:ODE_system_2} with the results of the numerical simulations~\citep{Kuhlen:2022tov}.} 
\label{fig:dFL_ODE} 
\end{figure}

We have solved this system numerically for $\eta = 0.2$ and $0.5$ and we show the resulting $d_{\text{FL}}(z)$ in Fig.~\ref{fig:dFL_ODE}. 
Note that we have rescaled the results from the numerical solution of eqs.~\eqref{eqn:ODE_system_1} and \eqref{eqn:ODE_system_2} by a factor $0.8$ and $0.56$ for $\eta = 0.2$ and $0.5$, respectively, to achieve agreement for small $z$. 
Overall, the agreement between simulations and this semi-analytical approach is rather good. 
At $z \ll \Lc$, field lines move ballistically and so $d_{\text{FL}}$ increases linearly with $z$. 
At $z \gg \Lc$, field lines travel diffusively such that $d_{\text{FL}} = \text{const.}$ 
In between, there is no direct transition; instead, $d_{\text{FL}}$ seems to overshoot and then oscillates before approaching the asymptotic value. 
These oscillation are seen both for the simulations and the semi-analytical approach, even though the period and phase do not match perfectly. 

Given the non-linear nature of the system of ODEs, it is difficult to understand the origin of the subdiffusive phase analytically.
However, in very broad terms, the three different phases can be understood qualitatively based on eqs.~\eqref{eqn:ODE_system_1} and \eqref{eqn:ODE_system_2}:
For small $z$, $\langle (\Delta x^{\text{FL}}(z))^2 \rangle$ is very small, such that both the exponential factor and the cosine in the integrand of eq.~\eqref{eqn:ODE_system_2} evaluate to $1$. 
The right hand side of eq.~\eqref{eqn:ODE_system_2} is then independent of $z$, such that $d_{\text{FL}} \propto z$. 
For very large $z$, the exponential term in eq.~\eqref{eqn:ODE_system_2} drives the the integrand to zero, such that  $d_{\text{FL}}$ becomes constant. 
The overshooting at intermediate times is due to the cosine term which changes sign. 

Intuitively, one might suspect that the overshooting in the running field-line diffusion coefficient is due to the sharp cut-off in the turbulence power spectrum for $k< k_0$, much like the Gibbs phenomenon~\citep{Bocher:1906} known from the Fourier transforms of functions with discontinuities. 
We have empirically confirmed this with the ODE approach by adjusting the power spectrum from a power law with cut-off for $k < k_0$ to a broken power law with $P(k) = \text{const.}$ for $k < k_0$. 
While the ringing is reduced, the subdiffusive behaviour persists. 

We note again that the semi-analytical results of $d_{\rm FL}$ can only match the simulational ones when rescaled by a factor decreasing with large turbulence level. This seems to suggest that the semi-analytical approach might be more appropriate for small turbulence level due to the simplified assumptions made at in deriving eqs. \eqref{eqn:ODE_system_1} and \eqref{eqn:ODE_system_2}. For this reason, we will not use the semi-analytical $d_{\rm FL}$ in the following but rather adopt a simple fit for the $d_{\rm FL}$ from simulations for each value of $\eta$ which takes into account the sub-diffusive phase of $d_{\rm FL}$ at the transition between ballistic and diffusive field line transport as observed for both semi-analytical and simulational results.

% ----------------------------------------------------------------------------------------
\subsubsection{The asymptotic perpendicular diffusion coefficient}

It is easy to see that eq.~\eqref{eqn:composition_dperp} will result in subdiffusive behaviour for the running perpendicular diffusion coefficient $d_{\perp}(t) \propto t^{-1/2}$ for times $t \gg \tau_{\text{d}}$ where particle transport in the parallel direction is diffusive, $\langle (\Delta z(t))^2 \rangle \propto t \Leftrightarrow d_{\parallel} = \kappa_{\parallel}$ and $z > \Lc$ such that field line transport is also diffusive, that is $d_{\text{FL}}(z) = \kappa_{\text{FL}}$. Here, $\kappa_{\parallel}$ and $\kappa_{\text{FL}}$ denote the $t \to \infty$ and $z \to \infty$ limits of $d_{\parallel}$ and $d_{\text{FL}}$, respectively. Note, however, that eq.~\eqref{eqn:composition_dperp} has been derived assuming that particles are attached to the field lines such that their perpendicular displacement is the same as the one for the field line (see eq. \eqref{eq:attach}). Once particles have travelled more than $\sim \Lc$ in the perpendicular direction they experience the transverse complexity, decouple from individual field lines and their perpendicular motion becomes diffusive~\citep{2019ApJ...881L..27S}, with the asymptotic diffusion coefficient $d_\perp(t>\tau_c)=d_{\perp}(\tau_c)= \kappa_{\perp} = \text{const.}$ The time $\tau_{\text{c}}$ is implicitly defined through 
\begin{equation}
d_{\perp}(\tau_{\text{c}}) \!=\! \frac{1}{2} \frac{\dd}{\dd t} \langle (\Delta x(t))^2 \rangle \Big|_{t \to \tau_{\text{c}}} \mkern-18mu \sim \frac{\langle (\Delta x (\tau_{\text{c}}))^2 \rangle}{2 \tau_{\text{c}}} \sim \frac{\Lc^2}{2 \tauc} . \label{eqn:def_tauc}
\end{equation}

In order to evaluate $\kappa_{\perp} = d_{\perp}(\tau_c)$ with eqs.~\eqref{eqn:composition_dperp} and \eqref{eqn:def_tauc}, we need to specify the functional forms for the running parallel and field line diffusion coefficients, $d_{\parallel}$ and $d_{\text{FL}}$. We parametrise the time-dependence of $d_{\parallel}$ with a broken power law, 
\begin{equation}
d_{\parallel}(t) = \kappa_{\parallel} \left( 1 + \left( t/\tau_{\text{s}} \right)^{-1/s_d} \right)^{-s_d}
\end{equation}
where $\tau_{\text{s}} = 3 \kappa_{\parallel} / v^2$ and $s_d = 1/2$. At early times, $t \ll \tau_{\text{s}}$, parallel particle transport is ballistic, $d_{\parallel}(t) \simeq \kappa_{\parallel} t/\tau_{\text{s}} = v^2 t/3$; at late times, $t \gg \tau_{\text{s}}$, it turns diffusive, $d_{\parallel}(t) \simeq \kappa_{\parallel}$. 

We model the asymptotic diffusion coefficient $\kappa_{\parallel}$ as a broken power law in $\rg/\Lc$~\citep{2009ASSL..362.....S}, 
\begin{equation}
\kappa_{\parallel}(\rg/\Lc) \! = \! A (\rg/\Lc)^{1/3} \left( 1 + \left( \frac{\rg/\Lc}{\rho_*} \right)^{5/(3 s_\kappa)} \right)^{s_\kappa}. \label{eqn:kappa_rho}
\end{equation}
At small rigidities, $\rg/\Lc \ll \rho_*$, CRs diffuse due to resonant interactions with turbulent fields and the rigidity-scaling $\kappa_{\parallel} \propto (\rg/\Lc)^{1/3}$ is characteristic for the Kolmogorov turbulence assumed. At large rigidities, $\rg/\Lc \gg \rho_*$, no resonant interactions are possible; instead, particles diffuse in the small-angle scattering limit $\kappa_{\parallel} \propto (\rg/\Lc)^2$~\citep{Gruzinov:2018yxz}. For each turbulence level $\eta$, there are thus three free parameters for $d_{\parallel}$: $A$, $\rho_*$, and $s_{\kappa}$ which are determined by fitting eq.~\eqref{eqn:kappa_rho} to the simulated $\kappa_{\parallel} = (3/v) \lambda_{\parallel}$. 
The fitted parameters are given in Tab.~\ref{tbl1}. 

\begin{table}[tbh]
\centering
\begin{tabular}{c | ccc }
$\eta$ & $A$ & $\rho_*$ & $s_{\kappa}$ \\
\hline
0.2 & 3.81 & 0.755 & 0.796 \\
0.5 & 1.06 & 0.627 & 0.906 \\
0.8 & 0.331 & 0.428 & 1.22 
\end{tabular}
\caption{
Model parameters for the rigidity-dependence of the asymptotic parallel diffusion coefficient $\kappa_{\parallel}$, see eq.~\eqref{eqn:kappa_rho}, for different turbulence levels $\eta$. 
}
\label{tbl1}
\end{table}

As far as $d_{\text{FL}}$ is concerned, our simulations have shown that field-line transport does not transition directly from ballistic to diffusive behaviour~\citep{Kuhlen:2022tov}. 
Instead, there is a subdiffusive phase where $d_{\text{FL}}$ decreases with time. 

We model the transition from ballistic to diffusive field line transport including an intermediate subdiffusive phase with a double-broken power law:
\begin{widetext}
\begin{equation}
d_{\text{FL}}(z) = C \, z \frac{\delta B_x^2}{B_0^2} \left(1 + \left( \frac{z}{z_1} \right)^{(1-\gamma)/s_1} \right)^{-s_1} \left(1 + \left( \frac{z}{z_2} \right)^{-\gamma/s_2} \right)^{s_2} \, . \label{eqn:dFL_parametrisation}
\end{equation}
\end{widetext}
Here, $C$ is a normalisation constant, $z_{1,2}$ are the distances along the field-line where the power law scaling of $d_{\text{FL}}$ changes and $s_{1,2}$ are the softnesses of these breaks. 
For small distances, $z \ll z_1$, field-line transport is ballistic, $d_{\text{FL}}(z) \simeq z (\delta B_x^2 / B_0^2)$ in line with analytical models of ballistic field line transport~\citep{2020SSRv..216...23S}. 
For $z_1 \ll z \ll z_2$, our parametrisation gives $d_{\text{FL}}(z) \propto z^{\gamma}$, thus allowing field line transport to be sub-diffusive, if $\gamma < 0$. 
For $z \gg z_2$, $d_{\text{FL}}(z) = \text{const.}$ 

\begin{figure}[thb]
\includegraphics[scale=1]{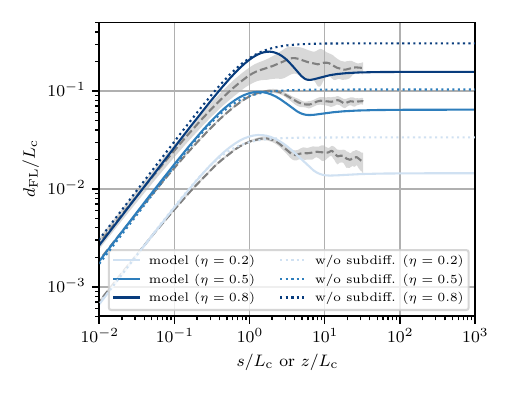}
\caption{
Running field line diffusion coefficients $d_{\text{FL}}$ for the different turbulence levels $\eta$. The solid lines show the running field-line diffusion coefficient $d_{\text{FL}}$ (eq.~\eqref{eqn:dFL_parametrisation}) including a subdiffusive phase with parameters fitted to the asymptotic perpendicular mean free path $\lambda_{\perp}$. For the dotted line, we have set $\gamma = 0$ in eq.~\eqref{eqn:dFL_parametrisation} and thus there is no subdiffusive phase. The grey dashed line and shaded band indicate the results of simulations for field line transport at the respective turbulence level $\eta$. 
}
\label{fig:d_FL}
\end{figure}

\begin{figure*}[!thb]
\includegraphics[scale=1]{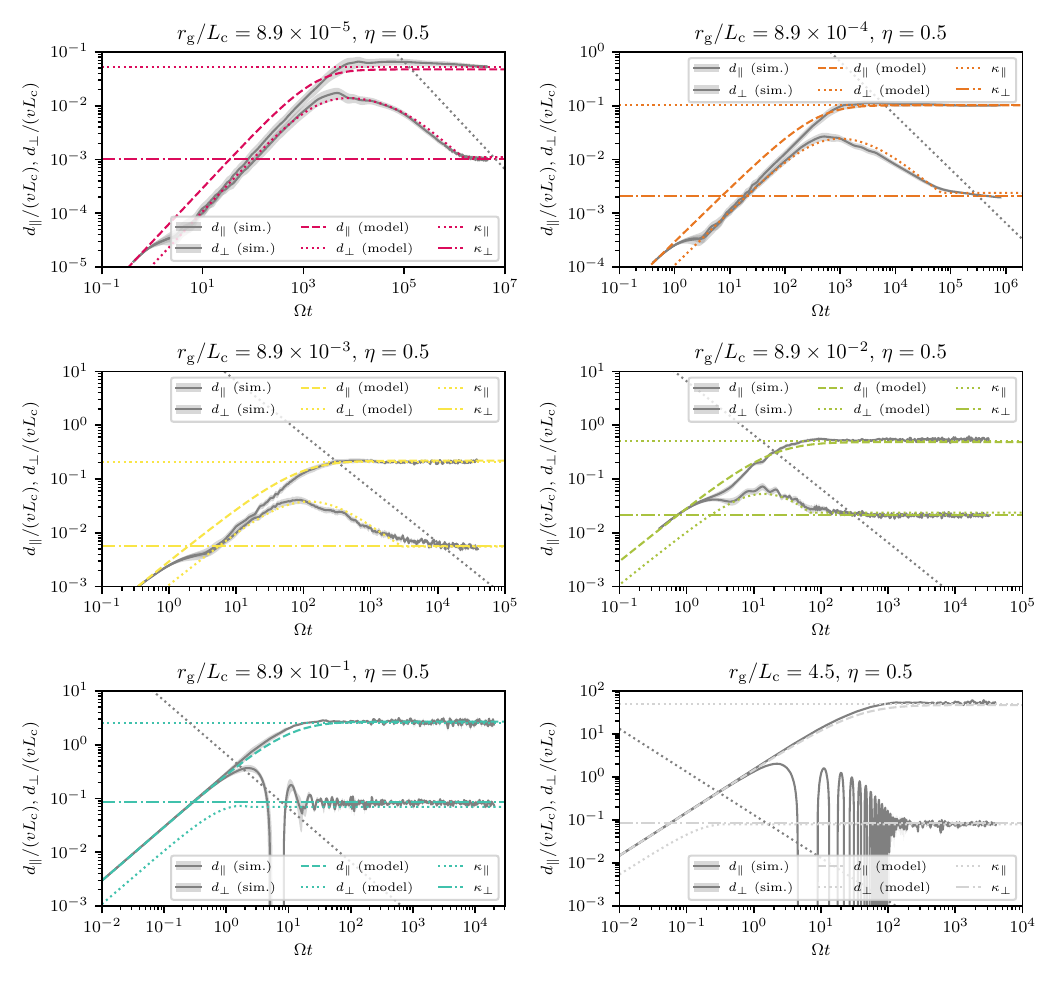}
\caption{
Running parallel and perpendicular diffusion coefficients $d_{\parallel}$ and $d_{\perp}$ for different reduced rigidities $\rg/\Lc$. The grey lines and bands indicate the means and standard mean errors from the test particle simulations. The dashed (solid) line shows the running parallel (perpendicular) diffusion coefficient from the simulations. The asymptotic values are indicated by the dashed and dot-dashed lines, respectively. Finally, the dotted grey line indicates $\Lc^2 / t$, that is the asymptotic perpendicular diffusion coefficient $\kappa_{\perp}$ is attained once the running perpendicular diffusion coefficient $d_{\perp}$ intersects this line.
}
\label{fig:dpar_dperp_rigidities}
\end{figure*}

By fitting to the simulated running field line diffusion coefficient for the three turbulence levels $\eta = 0.2, 0.5$ and $0.8$, we can fix all, but one of the parameters of eq.~\eqref{eqn:dFL_parametrisation}. 
Specifically, we find that $C = 0.8, z_1 = 1.5 \, \Lc$ and $z_2 = 8 \, \Lc$ reproduce the simulations well for $\eta = 0.2$. 
For $\eta = 0.5$, we use $C = 0.55, z_1 = 1.5 \, \Lc$ and $z_2 = 5 \, \Lc$ and for $\eta = 0.8$ we adopt $C = 0.2, z_1 = 2.5 \, \Lc$ and $z_2 = 5.5 \, \Lc$. 
In all cases, $s_1 = 1.5$ and $s_2 = 0.2$. 

In contrast, the decorrelation time $\tau_c$ is rather sensitive to the choice of $\gamma$ and so we have determined $\gamma$ by fitting to the simulated asymptotic perpendicular mean free path $\lambda_{\perp}$. 
Similarly, the above requirement for transverse complexity is only approximate. Therefore we replace $\Lc$ in eq.~\eqref{eqn:def_tauc} with another free parameter, $L_{\text{c},\perp}$. 
In summary, our model has two free parameters which cannot be fixed by fitting to the simulated running field line diffusion coefficient $d_{\text{FL}}$, but are determined by fitting our model prediction to the simulated asymptotic perpendicular mean free path $\lambda_{\perp}$. 
A comparison between the simulated running field line diffusion coefficient and our parametrisation thereof is shown in Fig.~\ref{fig:d_FL}.

% ----------------------------------------------------------------------------------------
% ----------------------------------------------------------------------------------------
% ----------------------------------------------------------------------------------------
\section{Results}

\begin{figure*}[thb]
\includegraphics[scale=1]{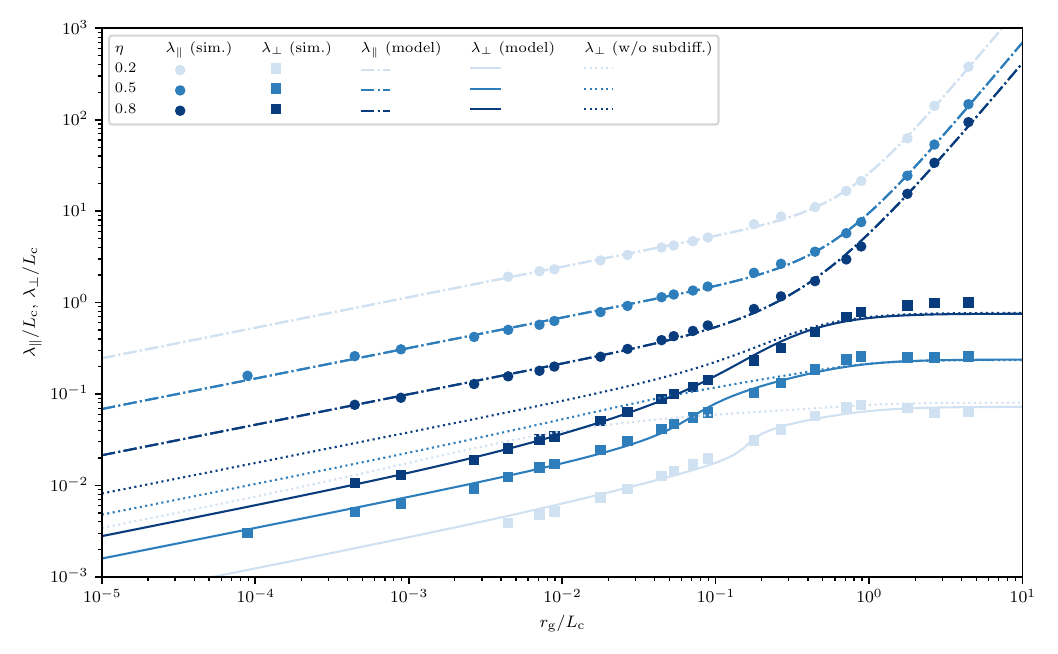}
\caption{Asymptotic parallel and perpendicular mean free paths $\lambda_{\parallel}$ and $\lambda_{\perp}$ as a function of reduced rigidity $\rg/\Lc$ for different turbulence levels $\eta$. The dots and squares are the results from test particle simulations. The dot-dashed and solid lines show the predictions from our model that includes a subdiffusive phase in the running field-line diffusion coefficient $d_{\text{FL}}$. The perpendicular mean-free path without a subdiffusive phase in $d_{\text{FL}}$ is indicated by the dotted lines.}
\label{fig:lams}
\end{figure*}

The first result of this paper are the parallel and perpendicular running diffusion coefficients. 
In Fig.~\ref{fig:dpar_dperp_rigidities} we compare the running diffusion coefficients from the test particle simulations with those of our model.

\begin{figure}[thb]
\includegraphics[width=\columnwidth]{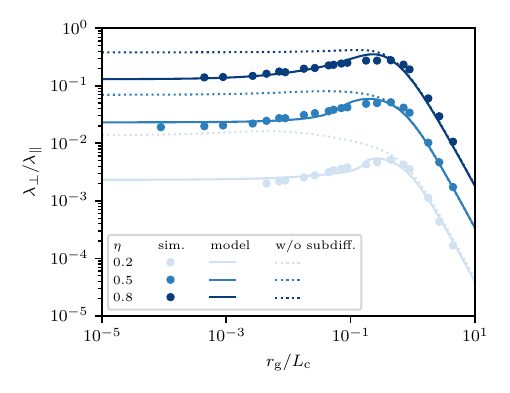}
\caption{Ratio $\lambda_{\perp} / \lambda{\parallel}$ of the perpendicular and parallel mean free paths as a function of reduced rigidity $\rg/\Lc$ for different turbulence levels $\eta$. The dots show the results of test particle simulations. Our model predictions including a subdiffusive phase in the running field-line diffusion coefficient $d_{\text{FL}}$ is shown by the solid lines. A model prediction without a subdiffusive phase in $d_{\text{FL}}$ is indicated by the dotted lines.}
\label{fig:ratio}
\end{figure}

The second result are the mean free paths, $\lambda_{\parallel}$ and $\lambda_{\perp}$ from our test particle simulations, shown in Fig.~\ref{fig:lams} by the dots and squares, respectively. The different shades of blue indicate different turbulence levels $\eta$. The parallel mean free path follows the form of eq.~\eqref{eqn:kappa_rho}, indicated by the dot-dashed lines. The perpendicular mean free path, however, exhibits a more complicated rigidity-dependence: At low rigidities, it has the same scaling as the parallel one, $\lambda_{\perp} \propto (\rg/\Lc)^{1/3}$, but at larger rigidities, the dependence is closer to $(\rg/\Lc)^{1/2}$. Eventually, for $\rg/\Lc \gg 1$, $\lambda_{\perp}$ becomes constant.  In Fig.~\ref{fig:ratio}, we also show the ratio of simulated perpendicular and parallel mean-free paths in order to better bring out the different rigidity scalings. 

Indications of the $(\rg/\Lc)^{1/2}$-scaling had been observed already earlier~\citep{2007JCAP...06..027D,2018JCAP...07..051G,2020PhRvD.102j3016D}, however, understanding of its origin was still lacking. Specifically, \citet{2020PhRvD.102j3016D} compared their simulation results with a few analytical theories~\citep{2003ApJ...590L..53M,2010ApJ...720L.127S} and concluded that none could reproduce the non-standard scaling at intermediate rigidities. To the best of our knowledge, we are the first to report the recovery of the standard scaling at lower rigidities. In the following, we will contrast our numerical results with those of our analytical model. 

In Fig.~\ref{fig:lams}, we also show the predictions of our analytical model, the second main result of this \textit{letter}. For the parallel mean free path, indicated by the dot-dashed line, this is the parametrisation of eq.~\eqref{eqn:kappa_rho} with the fitted parameters given in Tab.~\ref{tbl1}. For the perpendicular mean free path, indicted by the solid line, this is $\kappa_{\perp} = d_{\perp}(\tau_c)$ with eqs.~\eqref{eqn:composition_dperp} and \eqref{eqn:def_tauc}; the fitted parameters were given in the text following eq.~\eqref{eqn:dFL_parametrisation}. 

Overall, the model reproduces the mean free paths very well. In particular, the perpendicular mean free path $\lambda_{\perp}$ reproduces the complex rigidity-dependence found in the simulations. 

The cause of the non-standard rigidity-dependence for intermediate rigidities is the sub-diffusive phase in the running field-line diffusion coefficient. The fact that without such a subdiffusive phase, the non-standard rigidity-dependence for intermediate rigidities could not be reproduced is illustrated by the dotted lines in Figs.~\ref{fig:lams} and \ref{fig:ratio}. Here, we have assumed a running field line diffusion coefficient \emph{without} a subdiffusive phase, 
i.e.\ the same parameters for eq.~\eqref{eqn:dFL_parametrisation} as before but $\gamma = 0$.
With this, the $\lambda_{\perp}$ does not exhibit a non-standard rigidity-dependence for intermediate rigidities, but transitions almost smoothly between a $(\rg/\Lc)^{1/3}$-scaling and a constant.

% ----------------------------------------------------------------------------------------
% ----------------------------------------------------------------------------------------
% ----------------------------------------------------------------------------------------
\section{Conclusion}

To summarise, we have shown the results of a large suite of simulations of the transport of high-energy particles and field lines in isotropic Kolmogorov-like turbulence. The perpendicular diffusion coefficient in particular scales like $(\rg/\Lc)^{1/3}$ for low rigidities, $(\rg/\Lc)^{1/2}$ at intermediate rigidities and becomes constant for $\rg/\Lc \gg 1$. This behaviour is nicely reproduced in our analytical model and is ultimately due to a subdiffusive phase in the running field line diffusion coefficient. 

Our study has provided a number of crucial insights: First, we have confirmed the non-standard rigidity-scaling of the perpendicular mean free path at intermediate rigidities, but also found that it transitions back to the standard $(\rg/\Lc)^{1/3}$-dependence at the lowest rigidities. Second, we have for the first time provided a consistent analytical model that reproduces the non-standard behaviour. Our parametrised running and asymptotic diffusion coefficients are easy to implement in phenomenological studies or for comparison with future test particle simulations. Third, our model largely confirms the interpretation of perpendicular transport as due to particle transport along field lines and field line transport in the perpendicular direction, but with the need for a subdiffusive phase in the running field line diffusion coefficient. Finally, this understanding has given us renewed confidence in the standard scaling $\lambda_{\perp} \propto (\rg/\Lc)^{1/3}$ of the perpendicular diffusion coefficient at low rigidities. We provide a discussion of applications to a number of astrophysical problems in the companion paper~\citep{Kuhlen:2022tov}.

%\begin{acknowledgements}
The authors would like to thank Alexandre Marcowith for helpful discussions. This project was funded by the Deutsche Forschungsgemeinschaft (DFG, German Research Foundation) -- project number 426614101.
%\end{acknowledgements}

\bibliographystyle{aasjournal}
\bibliography{heuristic}

\begin{thebibliography}{}
\expandafter\ifx\csname natexlab\endcsname\relax\def\natexlab#1{#1}\fi
\providecommand{\url}[1]{\href{#1}{#1}}
\providecommand{\dodoi}[1]{doi:~\href{http://doi.org/#1}{\nolinkurl{#1}}}
\providecommand{\doeprint}[1]{\href{http://ascl.net/#1}{\nolinkurl{http://ascl.net/#1}}}
\providecommand{\doarXiv}[1]{\href{https://arxiv.org/abs/#1}{\nolinkurl{https://arxiv.org/abs/#1}}}

\bibitem[{{Bieber} \& {Matthaeus}(1997)}]{1997ApJ...485..655B}
{Bieber}, J.~W., \& {Matthaeus}, W.~H. 1997, \apj, 485, 655,
  \dodoi{10.1086/304464}

\bibitem[{Bôcher(1906)}]{Bocher:1906}
Bôcher, M. 1906, Annals of Mathematics, 7, 81.
\newblock \url{http://www.jstor.org/stable/1967238}

\bibitem[{{Casse} {et~al.}(2001){Casse}, {Lemoine}, \&
  {Pelletier}}]{2001PhRvD..65b3002C}
{Casse}, F., {Lemoine}, M., \& {Pelletier}, G. 2001, \prd, 65, 023002,
  \dodoi{10.1103/PhysRevD.65.023002}

\bibitem[{{Cerri} {et~al.}(2017){Cerri}, {Gaggero}, {Vittino}, {Evoli}, \&
  {Grasso}}]{2017JCAP...10..019C}
{Cerri}, S.~S., {Gaggero}, D., {Vittino}, A., {Evoli}, C., \& {Grasso}, D.
  2017, \jcap, 2017, 019, \dodoi{10.1088/1475-7516/2017/10/019}

\bibitem[{{Corrsin}(1959)}]{1959AdGeo...6..161C}
{Corrsin}, S. 1959, Advances in Geophysics, 6, 161,
  \dodoi{10.1016/S0065-2687(08)60102-8}

\bibitem[{{DeMarco} {et~al.}(2007){DeMarco}, {Blasi}, \&
  {Stanev}}]{2007JCAP...06..027D}
{DeMarco}, D., {Blasi}, P., \& {Stanev}, T. 2007, \jcap, 2007, 027,
  \dodoi{10.1088/1475-7516/2007/06/027}

\bibitem[{{Dundovic} {et~al.}(2020){Dundovic}, {Pezzi}, {Blasi}, {Evoli}, \&
  {Matthaeus}}]{2020PhRvD.102j3016D}
{Dundovic}, A., {Pezzi}, O., {Blasi}, P., {Evoli}, C., \& {Matthaeus}, W.~H.
  2020, \prd, 102, 103016, \dodoi{10.1103/PhysRevD.102.103016}

\bibitem[{{Evoli} {et~al.}(2012){Evoli}, {Gaggero}, {Grasso}, \&
  {Maccione}}]{2012PhRvL.108u1102E}
{Evoli}, C., {Gaggero}, D., {Grasso}, D., \& {Maccione}, L. 2012, \prl, 108,
  211102, \dodoi{10.1103/PhysRevLett.108.211102}

\bibitem[{{Giacalone} \& {Jokipii}(1999)}]{1999ApJ...520..204G}
{Giacalone}, J., \& {Jokipii}, J.~R. 1999, \apj, 520, 204,
  \dodoi{10.1086/307452}

\bibitem[{{Giacinti} {et~al.}(2018){Giacinti}, {Kachelrie\ss{}}, \&
  {Semikoz}}]{2018JCAP...07..051G}
{Giacinti}, G., {Kachelrie\ss{}}, M., \& {Semikoz}, D.~V. 2018, \jcap, 2018,
  051, \dodoi{10.1088/1475-7516/2018/07/051}

\bibitem[{{Giacinti} {et~al.}(2012){Giacinti}, {Kachelrie{\ss}}, {Semikoz}, \&
  {Sigl}}]{2012JCAP...07..031G}
{Giacinti}, G., {Kachelrie{\ss}}, M., {Semikoz}, D.~V., \& {Sigl}, G. 2012,
  \jcap, 2012, 031, \dodoi{10.1088/1475-7516/2012/07/031}

\bibitem[{Gruzinov(2018)}]{Gruzinov:2018yxz}
Gruzinov, A. 2018.
\newblock \doarXiv{1808.00041}

\bibitem[{{Hall} \& {Sturrock}(1967)}]{1967PhFl...10.2620H}
{Hall}, D.~E., \& {Sturrock}, P.~A. 1967, Physics of Fluids, 10, 2620,
  \dodoi{10.1063/1.1762084}

\bibitem[{{Harari} {et~al.}(2002){Harari}, {Mollerach}, {Roulet}, \&
  {S{\'a}nchez}}]{2002JHEP...03..045H}
{Harari}, D., {Mollerach}, S., {Roulet}, E., \& {S{\'a}nchez}, F. 2002, Journal
  of High Energy Physics, 2002, 045, \dodoi{10.1088/1126-6708/2002/03/045}

\bibitem[{{Hasselmann} \& {Wibberenz}(1970)}]{1970ApJ...162.1049H}
{Hasselmann}, K., \& {Wibberenz}, G. 1970, \apj, 162, 1049,
  \dodoi{10.1086/150736}

\bibitem[{{Jokipii}(1966)}]{1966ApJ...146..480J}
{Jokipii}, J.~R. 1966, \apj, 146, 480, \dodoi{10.1086/148912}

\bibitem[{{Kennel} \& {Engelmann}(1966)}]{1966PhFl....9.2377K}
{Kennel}, C.~F., \& {Engelmann}, F. 1966, Physics of Fluids, 9, 2377,
  \dodoi{10.1063/1.1761629}

\bibitem[{{Kolmogorov}(1941)}]{1941DoSSR..30..301K}
{Kolmogorov}, A. 1941, Akademiia Nauk SSSR Doklady, 30, 301

\bibitem[{{Kubo}(1962)}]{1962JPSJ...17.1100K}
{Kubo}, R. 1962, Journal of the Physical Society of Japan, 17, 1100,
  \dodoi{10.1143/JPSJ.17.1100}

\bibitem[{{Kuhlen} {et~al.}(2022){Kuhlen}, {Phan}, \&
  {Mertsch}}]{2022ApJ...927..110K}
{Kuhlen}, M., {Phan}, V. H.~M., \& {Mertsch}, P. 2022, \apj, 927, 110,
  \dodoi{10.3847/1538-4357/ac503b}

\bibitem[{Kuhlen {et~al.}(2022)Kuhlen, Phan, \& Mertsch}]{Kuhlen:2022tov}
Kuhlen, M., Phan, V. H.~M., \& Mertsch, P. 2022.
\newblock \doarXiv{2211.05881}

\bibitem[{{Lasuik} \& {Shalchi}(2017)}]{2017ApJ...847....9L}
{Lasuik}, J., \& {Shalchi}, A. 2017, \apj, 847, 9,
  \dodoi{10.3847/1538-4357/aa8720}

\bibitem[{{Matthaeus} {et~al.}(2003){Matthaeus}, {Qin}, {Bieber}, \&
  {Zank}}]{2003ApJ...590L..53M}
{Matthaeus}, W.~H., {Qin}, G., {Bieber}, J.~W., \& {Zank}, G.~P. 2003, \apjl,
  590, L53, \dodoi{10.1086/376613}

\bibitem[{{Mertsch}(2020)}]{2020Ap&SS.365..135M}
{Mertsch}, P. 2020, \apss, 365, 135, \dodoi{10.1007/s10509-020-03832-3}

\bibitem[{{Pucci} {et~al.}(2016){Pucci}, {Malara}, {Perri}, {Zimbardo},
  {Sorriso-Valvo}, \& {Valentini}}]{2016MNRAS.459.3395P}
{Pucci}, F., {Malara}, F., {Perri}, S., {et~al.} 2016, \mnras, 459, 3395,
  \dodoi{10.1093/mnras/stw877}

\bibitem[{{Qin} {et~al.}(2002{\natexlab{a}}){Qin}, {Matthaeus}, \&
  {Bieber}}]{2002GeoRL..29.1048Q}
{Qin}, G., {Matthaeus}, W.~H., \& {Bieber}, J.~W. 2002{\natexlab{a}}, \grl, 29,
  1048, \dodoi{10.1029/2001GL014035}

\bibitem[{{Qin} {et~al.}(2002{\natexlab{b}}){Qin}, {Matthaeus}, \&
  {Bieber}}]{2002ApJ...578L.117Q}
---. 2002{\natexlab{b}}, \apjl, 578, L117, \dodoi{10.1086/344687}

\bibitem[{{Reichherzer} {et~al.}(2020){Reichherzer}, {Becker Tjus}, {Zweibel},
  {Merten}, \& {Pueschel}}]{2020MNRAS.498.5051R}
{Reichherzer}, P., {Becker Tjus}, J., {Zweibel}, E.~G., {Merten}, L., \&
  {Pueschel}, M.~J. 2020, \mnras, 498, 5051, \dodoi{10.1093/mnras/staa2533}

\bibitem[{{Reichherzer} {et~al.}(2022){Reichherzer}, {Merten}, {D{\"o}rner},
  {Becker Tjus}, {Pueschel}, \& {Zweibel}}]{2022SNAS....4...15R}
{Reichherzer}, P., {Merten}, L., {D{\"o}rner}, J., {et~al.} 2022, SN Applied
  Sciences, 4, 15, \dodoi{10.1007/s42452-021-04891-z}

\bibitem[{{Schlegel} {et~al.}(2020){Schlegel}, {Frie}, {Eichmann},
  {Reichherzer}, \& {Tjus}}]{2020ApJ...889..123S}
{Schlegel}, L., {Frie}, A., {Eichmann}, B., {Reichherzer}, P., \& {Tjus}, J.~B.
  2020, \apj, 889, 123, \dodoi{10.3847/1538-4357/ab643b}

\bibitem[{{Shalchi}(2009)}]{2009ASSL..362.....S}
{Shalchi}, A. 2009, {Nonlinear Cosmic Ray Diffusion Theories}, Vol. 362
  (Springer Berlin, Heidelberg), \dodoi{10.1007/978-3-642-00309-7}

\bibitem[{{Shalchi}(2010)}]{2010ApJ...720L.127S}
---. 2010, \apjl, 720, L127, \dodoi{10.1088/2041-8205/720/2/L127}

\bibitem[{{Shalchi}(2017)}]{2017PhPl...24e0702S}
---. 2017, Physics of Plasmas, 24, 050702, \dodoi{10.1063/1.4982805}

\bibitem[{{Shalchi}(2019)}]{2019ApJ...881L..27S}
---. 2019, \apjl, 881, L27, \dodoi{10.3847/2041-8213/ab379d}

\bibitem[{{Shalchi}(2020)}]{2020SSRv..216...23S}
---. 2020, \ssr, 216, 23, \dodoi{10.1007/s11214-020-0644-4}

\bibitem[{{Shalchi} {et~al.}(2004){Shalchi}, {Bieber}, {Matthaeus}, \&
  {Qin}}]{2004ApJ...616..617S}
{Shalchi}, A., {Bieber}, J.~W., {Matthaeus}, W.~H., \& {Qin}, G. 2004, \apj,
  616, 617, \dodoi{10.1086/424839}

\bibitem[{{Snodin} {et~al.}(2016){Snodin}, {Shukurov}, {Sarson}, {Bushby}, \&
  {Rodrigues}}]{2016MNRAS.457.3975S}
{Snodin}, A.~P., {Shukurov}, A., {Sarson}, G.~R., {Bushby}, P.~J., \&
  {Rodrigues}, L.~F.~S. 2016, \mnras, 457, 3975, \dodoi{10.1093/mnras/stw217}

\bibitem[{{Sonsrettee} {et~al.}(2016){Sonsrettee}, {Subedi}, {Ruffolo},
  {Matthaeus}, {Snodin}, {Wongpan}, {Chuychai}, {Rowlands}, \&
  {Vyas}}]{2016ApJS..225...20S}
{Sonsrettee}, W., {Subedi}, P., {Ruffolo}, D., {et~al.} 2016, \apjs, 225, 20,
  \dodoi{10.3847/0067-0049/225/2/20}

\bibitem[{{Subedi} {et~al.}(2017){Subedi}, {Sonsrettee}, {Blasi}, {Ruffolo},
  {Matthaeus}, {Montgomery}, {Chuychai}, {Dmitruk}, {Wan}, {Parashar}, \&
  {Chhiber}}]{2017ApJ...837..140S}
{Subedi}, P., {Sonsrettee}, W., {Blasi}, P., {et~al.} 2017, \apj, 837, 140,
  \dodoi{10.3847/1538-4357/aa603a}

\bibitem[{{Tautz}(2016)}]{2016NewA...45....1T}
{Tautz}, R.~C. 2016, \na, 45, 1, \dodoi{10.1016/j.newast.2015.10.012}

\end{thebibliography}

\appendix
% ----------------------------------------------------------------------------------------
% ----------------------------------------------------------------------------------------
% ----------------------------------------------------------------------------------------
% ----------------------------------------------------------------------------------------
\section{Microscopic derivation for running perpendicular diffusion coefficient}
\label{appendix:micro}

In the following, we provide a microscopic derivation for eq.~\eqref{eqn:composition_dperp}, which bears some resemblance with time-dependent unified non-linear theory (UNLT)~\citep{2017PhPl...24e0702S,2017ApJ...847....9L}. 
Note however, that time-dependent UNLT has not been applied to isotropic turbulence before. 
We start from the assumption that up to the time where transverse complexity becomes important, particles follow field lines,
\begin{equation}
v_x = \frac{\delta B_x}{B_0} v_z \, .
\end{equation}

The velocity autocorrelation therefore evaluates to 
\begin{align}
\langle v_x(t) v_x(0) \rangle &= \frac{1}{B_0^2} \langle \delta B_x(t) \delta B_x(0) v_z(t) v_z(0) \rangle \nonumber
\end{align}
Usually, this is evaluated by first assuming that the four point function factorises, e.g.~\citep{2003ApJ...590L..53M,2020SSRv..216...23S}, 
\begin{equation}
\langle \delta B_x(t) \delta B_x(0) v_z(t) v_z(0) \rangle \simeq \langle \delta B_x(t) \delta B_x(0) \rangle \langle v_z(t) v_z(0) \rangle \, , \label{eqn:four_to_two_point_function}
\end{equation}
and then using the Fourier representation for the magnetic field two-point function,
\begin{align}
& \langle \delta B_x(t) \delta B_x(0) \rangle \nonumber \\
&= \Big\langle \int \dd^3 k \!\! \int \!\! \dd^3 k' \, \ee^{i (\vct{k} \cdot \vct{x}(t) - \vct{k}' \cdot \vct{x}(0))} \delta \tilde{B}_x(\vct{k}) \delta \tilde{B}^*_x(\vct{k}') \Big\rangle \\
&= \int \dd^3 k \, \langle \ee^{i \vct{k} \cdot (\vct{x}(t) - \vct{x}(0))} \delta \tilde{B}_x(\vct{k}) \delta \tilde{B}^*_x(\vct{k}) \rangle \, . 
\end{align}
Applying Corrsin's independence hypothesis~\citep{1959AdGeo...6..161C}, this results in 
\begin{align}
\langle \delta B_x(t) \delta B_x(0) \rangle &= \int \dd^3 k \, \langle \ee^{i \vct{k} \cdot (\vct{x}(t) - \vct{x}(0))} \rangle \langle \delta \tilde{B}_x(\vct{k}) \delta \tilde{B}^*_x(\vct{k}) \rangle \, , \nonumber
\end{align}
which can be evaluated by making an assumption for the spatial distribution of particles, e.g. assuming it to be Gaussian. However, the approximation of eq.~\eqref{eqn:four_to_two_point_function} ignores correlations between particle positions and velocities and it has been shown~\citep{2020SSRv..216...23S} that this loss of correlations leads to erroneous results. For instance in slab turbulence, perpendicular particle transport is known to be subdiffusive, $\langle (\Delta x)^2 \rangle \propto t^{1/2}$ while the result following from the assumption of eq.~\eqref{eqn:four_to_two_point_function} is diffusive, $\langle (\Delta x)^2 \rangle \propto t$. Allowing for correlations between particle positions and velocities, however, perpendicular subdiffusion is recovered~\citep{2020SSRv..216...23S}. 

Here, we factorise the correlation function into terms which depend on the parallel transport, on the magentic field and on the perpendicular position, 
\begin{align}
& \langle v_x(t) v_x(0) \rangle \\
&= \frac{1}{B_0^2} \int \dd^3 k \langle v_z(t) v_z(0) \ee^{i \vct{k} \cdot \vct{x}} \delta \tilde{B}_x(\vct{k}) \delta \tilde{B}^*_x(\vct{k}) \rangle \nonumber \\
&\simeq \frac{1}{B_0^2} \int \dd^3 k \langle v_z(t) v_z(0) \ee^{i k_\parallel z} \rangle \langle \delta \tilde{B}_x(\vct{k}) \delta \tilde{B}^*_x(\vct{k}) \rangle \langle \ee^{i \vct{k}_{\perp} \cdot \vct{x}_{\perp}} \rangle \, . \label{eqn:perp_vel_autocorr}
\end{align}

% ----------------------------------------------------------------------------------------
\paragraph{Cumulant expansion}

In the following, we evaluate the first term in the integrand, $\langle v_z(t) v_z(0) \ee^{i k_\parallel z} \rangle$. With $v_z(t) = \dd z / \dd t$ we find
\begin{equation}
\langle v_z(t) v_z(0) \ee^{i k_\parallel z} \rangle = \frac{1}{i k_{\parallel}} \frac{\dd}{\dd t} \langle v_z(0) \ee^{i k_\parallel z} \rangle \, , \label{eqn:triple_correlation_function}
\end{equation}
and we expand the complex exponential into a power series,
\begin{equation}
\ee^{i k_\parallel z} = \sum_{n = 0}^{\infty} \frac{1}{n!} (i k_{\parallel} z)^n \, . \label{eqn:exponential_series}
\end{equation}

We assume Gaussian statistics, that is only correlation functions of even order (in $v_z$ and $z$) contribute and higher order correlation functions can be expressed in terms of products of two-point functions. Therefore, in the exponential series of eq.~\eqref{eqn:exponential_series} only the odd terms contribute,
\begin{align}
\langle v_z(0) \ee^{i k_\parallel z} \rangle &= \sum_{n = 0}^{\infty} \frac{1}{n!} (i k_{\parallel})^n \langle v_z(0) z^n \rangle \nonumber \\
&= \sum_{p = 0}^{\infty} \frac{1}{(2p+1)!} (i k_{\parallel})^{2p+1} \langle v_z(0) z^{2p+1} \rangle \label{eqn:series2} \, .
\end{align}

A cumulant expansion~\citep{1962JPSJ...17.1100K} gives,
\begin{align*}
& \langle v_z(0) z^{2p+1} \rangle \\
&= (2p+1) \langle v_z(0) z \rangle \langle z^{2p} \rangle \\
&= (2p+1) \langle v_z(0) z \rangle (2p-1)!! \langle z^2 \rangle^p \, .
\end{align*}
We can now exploit the fact that $(2p+1)!! / (2p+1)! = 1/2^p/(p!)$ and perform the sum in eq.~\eqref{eqn:series2},
\begin{align}
& \langle v_z(0) \ee^{i k_\parallel z} \rangle \nonumber \\
&= \sum_{p = 0}^{\infty} \frac{1}{(2p+1)!} (i k_{\parallel})^{2p+1} (2p+1) \langle v_z(0) z \rangle (2p-1)!! \langle z^2 \rangle^p \nonumber \\
&= i k_{\parallel} \sum_{p = 0}^{\infty} \frac{(2p+1)!!}{(2p+1)!} (i k_{\parallel})^{2p} \langle v_z(0) z \rangle \langle z^2 \rangle^p \nonumber \\
&= i k_{\parallel} \langle v_z(0) z \rangle \sum_{p = 0}^{\infty} \frac{1}{p!} \left( - \frac{1}{2} k^2_{\parallel} \langle z^2 \rangle \right)^{p} \nonumber \\
&= i k_{\parallel} \langle v_z(0) z \rangle \ee^{ -\frac{1}{2} k^2_{\parallel} \langle z^2 \rangle } \nonumber \\
&= i k_{\parallel} d_{\parallel}(t) \ee^{ -\frac{1}{2} k^2_{\parallel} \langle z^2 \rangle } \, . \label{eqn:result_cumulant_expansion}
\end{align}
In the last line we have used the fact that $\langle v_z(0) z \rangle = d_{\parallel}$ which follows from
\begin{align*}
d_{ij}(t) &= \int_0^t \dd t' \langle v_i(t') v_j(0) \rangle = \left\langle \int_0^{x_i} \dd x'_i v_j(0) \right\rangle \\
&= \langle x_i(t) v_j(0) \rangle \, .
\end{align*}

Combining eqs.~\eqref{eqn:triple_correlation_function} and \eqref{eqn:result_cumulant_expansion} we thus find,
\begin{align}
& \langle v_z(t) v_z(0) \ee^{i k_\parallel z} \rangle \\
&= \frac{\dd}{\dd t} \left( d_{\parallel}(t) \ee^{ -\frac{1}{2} k^2_{\parallel} \langle z^2 \rangle } \right) \label{eqn:result_w_correlations} \\
&= \frac{\dd}{\dd t} \Big( d_{\parallel}(t) \Big) \ee^{ -\frac{1}{2} k^2_{\parallel} \langle z^2 \rangle } - k_{\parallel}^2 \left( d_{\parallel}(t) \right)^2 \ee^{ -\frac{1}{2} k^2_{\parallel} \langle z^2 \rangle } \, . \label{eqn:result_w_correlations_expanded}
\end{align}
Note how this differs from the result obtained if the correlations between $v_z$ and $z$ were ignored,
\begin{align}
\langle v_z(t) v_z(0) \rangle \langle \ee^{i k_\parallel z} \rangle &= \frac{\dd}{\dd t} \Big( d_{\parallel}(t) \Big) \ee^{ -\frac{1}{2} k^2_{\parallel} \langle z^2 \rangle } \, . \label{eqn:result_wo_correlations}
\end{align}
In this case, the diffusivity is usually overestimated. The last term in eq.~\eqref{eqn:result_w_correlations_expanded} can be considered a correction due to correlations between parallel velocities and parallel positions. It reduces the velocity autocorrelation function, thus allowing for instance for sub-diffusive behaviour for the case of slab turbulence.

Eq.~\eqref{eqn:result_w_correlations} can be further simplified by again using that $d_{\parallel}(t) = (1/2) \dd \langle z^2 \rangle / \dd t$,
\begin{equation}
\langle v_z(t) v_z(0) \ee^{i k_\parallel z} \rangle = \frac{1}{(i k_{\parallel})^2} \frac{\dd^2}{\dd t^2} \Big( \ee^{- \frac{1}{2} k_\parallel^2 \langle z^2 \rangle} \Big) \, . \label{eqn:xi}
\end{equation}

% ----------------------------------------------------------------------------------------
\paragraph{Perpendicular mean-square displacement}

We substitute eq.~\eqref{eqn:xi} into the perpendicular velocity autocorrelation, eq.~\eqref{eqn:perp_vel_autocorr}, 
\begin{align}
& \langle v_x(t) v_x(0) \rangle \\
&\simeq \frac{1}{B_0^2} \int \dd^3 k \langle v_z(t) v_z(0) \ee^{i k_\parallel z} \rangle \langle \delta B_x(\vct{k}) \delta B^*_x(\vct{k}) \rangle \langle \ee^{i \vct{k}_{\perp} \cdot \vct{x}_{\perp}} \rangle \nonumber \\
&\simeq \frac{1}{B_0^2} \int \dd^3 k \frac{1}{(i k_{\parallel})^2} \frac{\dd^2}{\dd t^2} \Big( \ee^{- \frac{1}{2} k_\parallel^2 \langle z^2 \rangle} \Big) \langle \delta B_x(\vct{k}) \delta B^*_x(\vct{k}) \rangle \langle \ee^{i \vct{k}_{\perp} \cdot \vct{x}_{\perp}} \rangle \, . \label{eqn:perp_vel_autocorr-2}
\end{align}

Both in the ballistic and diffusive regime, the displacement in the perpendicular direction is much slower than in the parallel one. We therefore consider $\langle \ee^{i \vct{k}_{\perp} \cdot \vct{x}_{\perp}} \rangle \simeq \text{const.}$ over the time-scales where the integral has support and so, we can pull the double time-derivative out of the integral,
\begin{align}
& \langle v_x(t) v_x(0) \rangle \\
&\simeq \frac{1}{B_0^2} \frac{\dd^2}{\dd t^2} \int \dd^3 k \frac{1}{(i k_{\parallel})^2} \ee^{- \frac{1}{2} k_\parallel^2 \langle z^2 \rangle} \langle \delta B_x(\vct{k}) \delta B^*_x(\vct{k}) \rangle \langle \ee^{i \vct{k}_{\perp} \cdot \vct{x}_{\perp}} \rangle \, . \label{eqn:perp_vel_autocorr-3}
\end{align}
Using the fact that 
\begin{equation}
\langle v_x(t) v_x(0) \rangle = \frac{\dd^2}{\dd t^2} \langle (\Delta x(t))^2 \rangle \, ,
\end{equation}
we can integrate eq.~\eqref{eqn:perp_vel_autocorr-3}, 
\begin{align}
\langle (\Delta x(t))^2 \rangle &\simeq \frac{2}{B_0^2} \int \dd^3 k \frac{1}{(i k_{\parallel})^2} \ee^{ -\frac{1}{2} k^2_{\parallel} \langle z^2 \rangle } \langle \delta B_x(\vct{k}) \delta B^*_x(\vct{k}) \rangle \langle \ee^{i \vct{k}_{\perp} \cdot \vct{x}_{\perp}} \rangle \nonumber \\
&= \frac{2}{B_0^2} \int \dd^3 k \frac{1}{(i k_{\parallel})^2} \int_{-\infty}^{\infty} \dd z \, \ee^{i k_{\parallel} z} \frac{1}{\sqrt{2 \pi \langle z^2 \rangle}} \ee^{-\frac{z^2}{2 \langle z^2 \rangle}} \langle \delta B_x(\vct{k}) \delta B^*_x(\vct{k}) \rangle \langle \ee^{i \vct{k}_{\perp} \cdot \vct{x}_{\perp}} \rangle \nonumber \\
&\simeq 2 \int_{-\infty}^{\infty} \dd z \, \frac{1}{B_0^2} \int \dd^3 k \frac{1}{(i k_{\parallel})^2} \ee^{i k_{\parallel} z} \langle \delta B_x(\vct{k}) \delta B^*_x(\vct{k}) \rangle \langle \ee^{i \vct{k}_{\perp} \cdot \vct{x}_{\perp}} \rangle \frac{1}{\sqrt{2 \pi \langle z^2 \rangle}} \ee^{-\frac{z^2}{2 \langle z^2 \rangle}} \\
&= 2 \int_{-\infty}^{\infty} \dd z \, \frac{1}{B_0^2} \int \dd^3 k \int \dd z \int \dd z \, \ee^{i k_{\parallel} z} \langle \delta B_x(\vct{k}) \delta B^*_x(\vct{k}) \rangle \langle \ee^{i \vct{k}_{\perp} \cdot \vct{x}_{\perp}} \rangle \frac{1}{\sqrt{2 \pi \langle z^2 \rangle}} \ee^{-\frac{z^2}{2 \langle z^2 \rangle}} \, . \label{eqn:perp_dist_autocorr}
\end{align}
Again, for small turbulence level, we evaluate the field line essentially along the $z$-axis and so we can approximate the Fourier transform
\begin{equation}
\int \dd^3 k \, \ee^{i k_{\parallel} z} \langle \delta B_x(\vct{k}) \delta B^*_x(\vct{k}) \rangle \langle \ee^{i \vct{k}_{\perp} \cdot \vct{x}_{\perp}} \rangle \simeq \langle \delta B_x(z) \delta B_x(0) \rangle \, .
\end{equation}
With this, eq.~\eqref{eqn:perp_dist_autocorr} gives
\begin{align}
\langle (\Delta x)^2 \rangle &\simeq 2 \int_{-\infty}^{\infty} \!\! \dd z \frac{1}{B_0^2} \int \!\! \dd z \! \int \!\! \dd z \langle \delta B_x(z) \delta B_x(0) \rangle \frac{\ee^{-\frac{z^2}{2 \langle z^2 \rangle}}}{\sqrt{2 \pi \langle z^2 \rangle}} \\
&= 2 \int_{-\infty}^{\infty} \dd z \, \frac{1}{\sqrt{2 \pi \langle z^2 \rangle}} \ee^{-\frac{z^2}{2 \langle z^2 \rangle}} \int_0^z \dd z' \, d_{\text{FL}}(z') \label{eqn:msd} \, .
\end{align}

Eq.~\eqref{eqn:msd} can be evaluated in two ways. 

% ----------------------------------------------------------------------------------------
\paragraph{Estimating the integral} 

This expression can be evaluated by again switching the order of integrations,
\begin{align}
\langle (\Delta x)^2 \rangle &= 4 \int_0^{\infty} \dd z' \, d_{\mathrm{FL}}(z') \int_{z'}^{\infty} \dd z \, \frac{1}{\sqrt{2 \pi \langle z^2 \rangle}} \ee^{-z^2/(2 \langle z^2 \rangle)} \nonumber \\
&= 2 \int_0^{\infty} \dd z' \, d_{\mathrm{FL}}(z') \mathrm{erfc} \left( \frac{z'}{\sqrt{2 \langle z^2 \rangle}} \right) \, ,
\label{eqn:msd-2}
\end{align}
where $\mathrm{erfc}(\cdot)$ denotes the complementary error function,
\begin{equation}
\mathrm{erfc}(\zeta) \equiv  \frac{2}{\sqrt{\pi}} \int_{\zeta}^{\infty} \dd \zeta' \,\ee^{-\zeta^2} \, .
\end{equation}
The factor $\mathrm{erfc} (z' / \sqrt{2 \langle z^2 \rangle})$ in the integrand of eq.~\eqref{eqn:msd-2} is suppressing contributions to the integral for $z' \gtrsim \sqrt{\langle z^2 \rangle / 2}$. Thus we can approximate the integral of eq.~\eqref{eqn:msd-2} as
\begin{equation}
\langle (\Delta x(t))^2 \rangle \simeq 2 \int_0^{\sqrt{\langle z^2 \rangle / 2}} \dd z' \, d_{\mathrm{FL}}(z') \, .
\end{equation}
Especially for the oftentimes found broken power law forms for $d_{\mathrm{FL}}(z')$ this is a good approximation for eq.~\eqref{eqn:msd-2}. Up to order one factors, this agrees with eq.~\eqref{eqn:rperp3} and so with eq.~\eqref{eqn:def_dperp}, again, we recover eq.~\eqref{eqn:composition_dperp}. 

\end{document}